\documentclass{ws-procs9x6}

\newcommand{\be}{\begin{eqnarray}} 
\newcommand{\ee}{\end{eqnarray}} 
\newcommand{\Msun}{\mbox{$M_{\odot}\;$}}

\begin{document}
%%%%%%%%%%%%%%%%%%%%%%%%%%%%%%%%%%%%%%%%%%%%%%%%%%%%%%%%%%%%%%%%%%%%%%%%
%%%%%%%%%%%%%%%%%%%%%%%%%%%%%%%%%%%%%%%%%%%%%%%%%%%%%%%%%%%%%%%%%%%%%%%%
%%%%%%%%%%%%%%%%%%%%%%%%%%%%%%%%%%%%%%%%%%%%%%%%%%%%%%%%%%%%%%%%%%%%%%%%

%%%%%%%%%%%%%%%%%%%%%%%%%%%%%%%%%%%%%%%%%%%%%%%%%%%%%%%%%%%%%%%%%%%%%%%%
%%%%%%%%%%%%%%%%%%%%%%%%%%%%%%%%%%%%%%%%%%%%%%%%%%%%%%%%%%%%%%%%%%%%%%%%
\title{The Minimal Cooling of Neutron Stars
       \footnote{\uppercase{W}ork in collaboration with
                 \uppercase{J.M. L}attimer, \uppercase{M. P}rakash, \&
                 \uppercase{A. W. S}teiner: 
                 \uppercase{R}ef. [1]}}
\author{Dany Page}
\address{Instituto de Astronom\'{\i}a, UNAM \\
         Mexico, D.F. 04340, MEXICO \\
         E-mail: page@astroscu.unam.mx}
\maketitle
%%%%%%%%%%%%%%%%%%%%%%%%%%%%%%%%%%%%%%%%%%%%%%%%%%%%%%%%%%%%%%%%%%%%%%%%
%%%%%%%%%%%%%%%%%%%%%%%%%%%%%%%%%%%%%%%%%%%%%%%%%%%%%%%%%%%%%%%%%%%%%%%%
\abstracts{
A general overview of the main physical processes driving the cooling of
an isolated neutron star is presented.
Among the most important ones are  the various possible neutrino 
emission processes and the occurrence of baryon pairing.
Special emphasis is also put on the importance of the chemical 
composition of the upper layers of the star.
A detailed analysis of a Minimal Scenario, which explicitly postulates 
that no ``exotic'' form of matter be present, is summarized and compared
with presently available observational data.
No striking incompatibility of the data with the predictions of the 
Minimal Scenario is found.
Nevertheless, two, possibly three, conspicuous stars are identified 
which may, when better data are available, constitute strong 
astrophysical evidence for the occurrence of a new state of matter at 
high density.}
%%%%%%%%%%%%%%%%%%%%%%%%%%%%%%%%%%%%%%%%%%%%%%%%%%%%%%%%%%%%%%%%%%%%%%%%
%%%%%%%%%%%%%%%%%%%%%%%%%%%%%%%%%%%%%%%%%%%%%%%%%%%%%%%%%%%%%%%%%%%%%%%%

%%%%%%%%%%%%%%%%%%%%%%%%%%%%%%%%%%%%%%%%%%%%%%%%%%%%%%%%%%%%%%%%%%%%%%%%
%%%%%%%%%%%%%%%%%%%%%%%%%%%%%%%%%%%%%%%%%%%%%%%%%%%%%%%%%%%%%%%%%%%%%%%%
\section{Introduction}
%%%%%%%%%%%%%%%%%%%%%%%%%%%%%%%%%%%%%%%%%%%%%%%%%%%%%%%%%%%%%%%%%%%%%%%%
%%%%%%%%%%%%%%%%%%%%%%%%%%%%%%%%%%%%%%%%%%%%%%%%%%%%%%%%%%%%%%%%%%%%%%%%

Among the various ways to search for new states of matter at high 
density the study of neutron stars is a promising one.
Many aspects of the very diverse phenomenology of these stars can 
provide us with indications of such ``exotic'' matter
(see, e.g., Ref.~[\refcite{vK04}]).
In particular, the modeling of the thermal evolution of isolated
neutron stars is an avenue along which much effort has been invested.
Being born in a supernova at temperatures in excess of 
$3\times 10^{11}$ K, young neutron stars rapidly cool through neutrino 
emission and the cooling rate is a very sensitive function of the 
composition of matter at the most extreme densities present in their
inner core.
Different models predict central densities from around 
$4 \times \rho_{\rm nucl}$ up to 15 to 20 times $\rho_{\rm nucl}$
($\rho_{\rm nucl}$ being the nuclear density), which may very probably
be within the necessary range to see deconfinement of baryonic matter 
into quark matter.
Less extreme models predict the occurrence of charged meson condensates
and/or hyperons populations.
However, distinguishing between these various scenarios is a very
delicate problem [\refcite{PPLS00}].
Finally, the most extreme model consider that neutron stars may convert
into ``Strange Stars'' made entirely of deconfined quark matter 
which would have a completely different thermal evolution
[\refcite{PU02}].

Evidence in favor of the presence of a new state of matter in the core 
of some neutron stars can only be obtained by finding some observed 
characteristics of these stars which cannot be understood  without the 
assumption of the presence of such matter.
Within this point of view I describe a ``Minimal Scenario'' of neutron 
star cooling, proposed recently in Ref.~[\refcite{PLPS04}], which 
precisely assumes that the neutron star interior is devoid of any form 
of matter beyond the standard composition consisting of only neutrons 
with a small admixture of protons, accompanied by the necessary amount
of electrons and muons to keep the star charge neutral.
This Minimal Scenario is a revised modern version of the ``Standard 
Scenario'' but incorporates, as an essential ingredient, the effects of
nucleon pairing, i.e., neutron superfluidity and/or proton 
superconductivity, on the star's specific heat and neutrino emission, 
particularly the neutrino emission by the very formation, and breaking,
of the Cooper pairs.
Comparison of the predictions of this Minimal Scenario with data may 
hence provide us with the long searched for evidence for ``exotic'' 
matter.

Section \ref{Sec:Data} briefly summarizes the presently available data 
on  temperature and luminosity  of isolated cooling neutron stars. 
Section \ref{Sec:Physics} describes the most important input physics for
the study of the Minimal Scenario.
Section \ref{Sec:Minimal} compares the results with data and 
Section \ref{Sec:Conclusion} offers conclusions.

An extensive presentation of this work can be found in 
Ref.~[\refcite{PLPS04}] to which the present summary could be considered
as a, hopefully convenient, {\em Traveler's Guide}.

%%%%%%%%%%%%%%%%%%%%%%%%%%%%%%%%%%%%%%%%%%%%%%%%%%%%%%%%%%%%%%%%%%%%%%%%
%%%%%%%%%%%%%%%%%%%%%%%%%%%%%%%%%%%%%%%%%%%%%%%%%%%%%%%%%%%%%%%%%%%%%%%%
\section{Observational Data
         \label{Sec:Data}}
%%%%%%%%%%%%%%%%%%%%%%%%%%%%%%%%%%%%%%%%%%%%%%%%%%%%%%%%%%%%%%%%%%%%%%%%
%%%%%%%%%%%%%%%%%%%%%%%%%%%%%%%%%%%%%%%%%%%%%%%%%%%%%%%%%%%%%%%%%%%%%%%%

%-----------------------------------------------------------------------
\begin{figure}[b]
   \centerline{\epsfxsize=4.5in\epsfbox{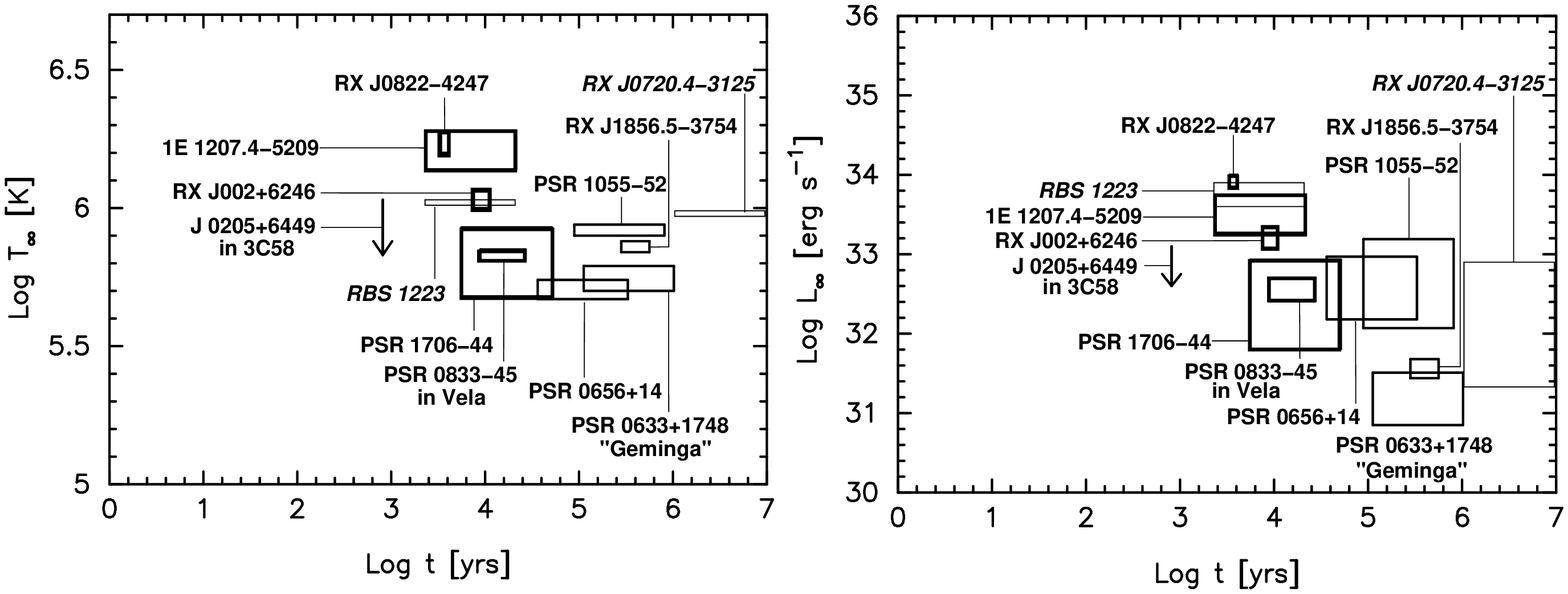}}   
   \caption{Measured $T_\infty$ and $L_\infty$, for twelve isolated 
            neutron stars, versus age. 
            The age, and their error bars, are either from
            kinematical information when available or from the pulsar 
            spin-down time scale, in which case an uncertainty of a 
            factor three has been assumed. 
            See [\protect\refcite{PLPS04}] for references and more
            details.
   \label{Fig:Data}}
\end{figure}
%-----------------------------------------------------------------------

Numerical calculations of neutron star cooling give as a natural result
the evolution of the star's photon thermal luminosity $L$ as a function 
of time.
This luminosity can equally well be described in terms of an effective
temperature $T_e$ through the standard relation
\be
L \equiv 4 \pi R^2 \cdot  \sigma_{\scriptscriptstyle S\! B} T_e^4
\;\;\;\;\;\;\;\; \mathrm{or} \;\;\;\;\;\;\;\;
L_\infty \equiv 4 \pi R_\infty^2 \cdot  
     \sigma_{\scriptscriptstyle S\! B} T_{e \, \infty}^4
\label{Eq:LTeff}
\ee
($\sigma_{\scriptscriptstyle S\! B}$ being the Stefan-Boltzmann 
constant) where $R$ is the star's radius and the subscripts $\infty$ 
indicate quantities as observed ``at infinity''.

Observations of cooling neutron stars can provide us with data in the 
form of luminosity $L_\infty$ and/or temperature $T_\infty$ at infinity.
The measured $T_\infty$ depends of course on the kind atmosphere assumed
in the spectral fits, realistic neutron star atmosphere models giving
generally lower values than blackbodies.
The measured $L_\infty$ is obtained from the total observed thermal 
flux, corrected for interstellar absorption, and the distance $D$.
If $D$ is known with sufficient accuracy Eq.~\ref{Eq:LTeff} could be 
used to determine $R_\infty$ [\refcite{D03}]
{\it assuming that $T_e$ is also accurately known}, i.e., that
the correct atmosphere model has been used in the spectral fit.
If the deduced $R_\infty$ is too small or too large compared to the 
``canonical'' 10 km expected for a neutron star it is a strong 
indication that the atmosphere model is not correct.
Nevertheless, some exotic models of compact stars as ``Strange Stars''
may result in small radii and also some magnetic field configurations
may be able to confine the detectable surface thermal emission
to an area significantly smaller than the whole surface of the star
[\refcite{GKP04}].

The data I will use are shown in Fig.~\ref{Fig:Data} and have been 
selected according to this self-consistency $R_\infty$-criterium.
Two types of spectra have been preferentially used in the spectral fits
producing these data: blackbodies and magnetized hydrogen atmospheres.
Only the second one has been successful in deducing acceptable values
for $R_\infty$ and this lead to the selection of $T_\infty$ and
$L_\infty$ of the objects plotted with thick lines in 
Fig.~\ref{Fig:Data}.
For the objects plotted with thin lines the magnetized hydrogen 
atmosphere models require much too large radii while blackbodies seem 
more reasonable but usually on the low side of the expected range of 
$R_\infty$.
Given this situation, for these objects it is difficult to decide which
of $T_\infty$ or  $L_\infty$ is the more reliable value to use for 
comparison with the theoretical models and I hence prefer to use both, 
leaving the reader draw her/his own conclusions from the analysis.

%%%%%%%%%%%%%%%%%%%%%%%%%%%%%%%%%%%%%%%%%%%%%%%%%%%%%%%%%%%%%%%%%%%%%%%%
%%%%%%%%%%%%%%%%%%%%%%%%%%%%%%%%%%%%%%%%%%%%%%%%%%%%%%%%%%%%%%%%%%%%%%%%
\section{The Physics of Neutron Star Cooling
         \label{Sec:Physics}}
%%%%%%%%%%%%%%%%%%%%%%%%%%%%%%%%%%%%%%%%%%%%%%%%%%%%%%%%%%%%%%%%%%%%%%%%
%%%%%%%%%%%%%%%%%%%%%%%%%%%%%%%%%%%%%%%%%%%%%%%%%%%%%%%%%%%%%%%%%%%%%%%%

%-----------------------------------------------------------------------
\begin{figure}[t]
   \centerline{\epsfxsize=3.0in\epsfbox{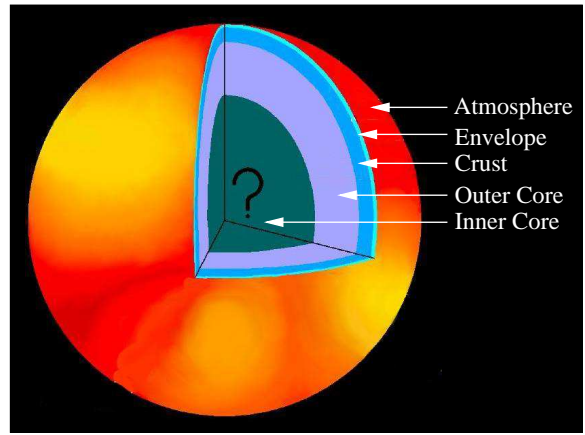}}   
   \caption{The different layers in a neutron star.
           [Drawing by the author.]
   \label{Fig:NStar}}
\end{figure}
%-----------------------------------------------------------------------

The picture in Fig.~\ref{Fig:NStar} illustrates the most important
layers in a neutron star: \\
- {\bf Atmosphere:} at most a few tens of centimeters thick, this is the
visible surface of the star (it may actually be a solid surface instead
of an atmosphere) where the thermal photons are emitted. 
It is of utmost importance for observations since it is where the 
energy distribution of  the thermally emitted photons, i.e., the thermal
{\em spectrum}, is determined.
However, since all the heat flowing into it from the interior is 
reemitted at the surface, the atmosphere does not affect the thermal 
evolution of the star.
\\
- {\bf Envelope:} this layer is several tens of meters thick and is,
by definition, where a large temperature gradient is always present. It
is a throttle which controls how much heat can leak out of the star and 
thus determines the relationship between the interior temperature and 
the effective temperature or, equivalently, the surface photon 
luminosity $L_\gamma$.
\\
- {\bf Crust:} with a thickness of several hundred meters, this layer
is important mostly in the cooling of very young stars or in the study
of transient phenomena as glitches. For our present purpose its only 
relevance is its (small) contribution to the specific heat.
\\
In both the envelope and the crust matter is made of nuclei immersed
in a gas of electrons and, in the inner part of the crust, at densities
higher than $\rho_{\rm drip} \simeq 4.3 \times 10^{11}$ gm cm$^{-3}$, 
a quantum liquid of dripped neutrons. \\
- {\bf Outer Core:} region at densities higher than
$\rho_{\rm cc} \simeq 1.6 \times 10^{14}$ gm cm$^{-3}$, where matter is
a quantum liquid predominantly composed of neutrons with a small 
fraction of protons, plus electrons and muons to maintain charge 
neutrality.
\\
- {\bf Inner Core:} the mysterious part, which may or may not exist,
and where ``exotic'' forms of matter may appear.
In the Minimal Scenario this inner core is explicitly assumed to be 
non-existent.

All calculations I will present here were performed with a wholly 
general relativistic Henyey-type stellar evolution code which solves 
exactly the equations of energy balance and heat transport inside a star
whose structure is calculated by solving the Tolman-Oppenheimer-Volkov
equation of hydrostatic equilibrium.
Nevertheless, the most important features can be understood from the
(Newtonian) energy conservation equation
\be
\frac{dE_{th}}{dt} = C_v \frac{dT}{dt}
                   = -L_\nu - L_\gamma + H
\label{Eq:energy-conservation}
\ee
where $E_{th}$ is the thermal energy content of the star, $L_\nu$
the neutrino luminosity, $L_\gamma$ the surface photon luminosity
and $H$ would give the contribution from ``heating processes'' as,
e.g., friction within the differentially rotating neutron superfluid or
magnetic field decay, and $C_v$ is the total specific heat,
$T$ being the interior temperature.
Solving the heat transport equation gives us the detailed temperature
profile in the interior but within a few tens of years after its birth
the star becomes isothermal and its evolution is then entirely 
controlled by (the GR version of) Eq.~\ref{Eq:energy-conservation}.
At this time a significant temperature gradient is only present in
the envelope (see, however, Ref.~[\refcite{GKP04}]).

%%%%%%%%%%%%%%%%%%%%%%%%%%%%%%%%%%%%%%%%%%%%%%%%%%%%%%%%%%%%%%%%%%%%%%%%
\subsection{The envelope and the photon luminosity
            \label{Sec:Envelope}}

\begin{figure}[ht]
   \centerline{\epsfxsize=2.5in\epsfbox{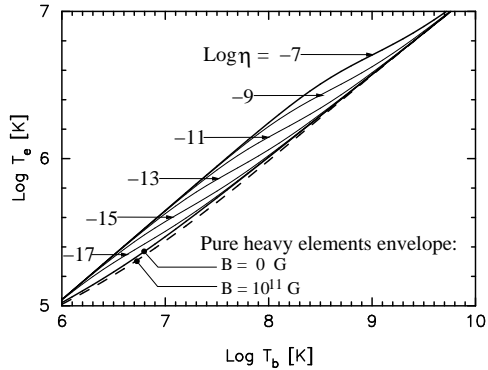}}   
   \caption{Relationship between the effective temperature $T_e$ and
            the interior temperature $T_b$ at the bottom of the
            envelope assuming various amounts of light elements
            parameterized by 
            $\eta \equiv g_{s \, 14}^2 \Delta M_{\rm L}/M$ where 
            $\Delta M_{\rm L}$ is the mass in light elements in the
            envelope and $g_{s \, 14}$ the surface gravity  in units of
            $10^{14}$ cm s$^{-1}$, $M$ being the total star's mass, in
            the absence of a magnetic field [\protect\refcite{PCY97}].
            Also shown are the $T_b - T_e$ relationships for an 
            envelope of heavy elements with and without the presence of
            a dipolar field of strength of $10^{11}$ G following 
            Ref.~[\protect\refcite{PY01}].}
   \label{Fig:Tb_Te} 
%\vspace{-0.1in}
\end{figure}

Once the star is isothermal its interior temperature is equal to the 
temperature at the bottom of the envelope, $T_b$, and the relationship
between $T_b$ and the ``surface'', or effective, temperature is 
called the ``$T_b$ -- $T_e$ relationship'',
which then gives us $L_\gamma$ through Eq.~\ref{Eq:LTeff}.
A useful approximation to it is [\refcite{GPE83}]
\be
T_e \; \sim \; \sqrt{T_b}
\;\;\;\;\;\;\;\; {\rm with} \;\;\;\;\;\;\;\; 
T_e \approx 10^6 \; {\rm K} \; \longleftrightarrow \;
 T_b \approx 10^8 \; {\rm K.}
\label{Eq:GPE}
\ee
which gives, very roughly, $L_\gamma \sim T^2$.
Nevertheless, significant deviation from Eq.~\ref{Eq:GPE} can occur.
This equation is based on models which assumed that no magnetic field
is present and that the envelope is made of iron, and iron-like, nuclei,
but in case light elements, e.g., $H$, $He$, $C$, or $O$, are present 
they strongly reduce the blanketing effect of the envelope.
A magnetic field also increases the heat permeability of the envelope
in the regions where it is pointing radially but strongly suppresses
it in regions where it makes small angles with the surface, thus
inducing a highly non-uniform surface temperature distribution
(see, e.g., Ref.~[\refcite{P95}]), and motivating the shaded surface in
Fig.~\ref{Fig:NStar}. 
Nevertheless the overall effect of the magnetic field is not as strong
as the effect of the chemical composition.
Figure~\ref{Fig:Tb_Te} shows this $T_b$ -- $T_e$ relationship for 
various models of envelope with varying amounts of light elements and 
an envelope formed entirely of heavy iron-like elements with and 
without a magnetic field.
Notice that an envelope with a significant amount of light elements 
results, for a given interior temperature $T_b$, in a luminosity 
$L_\gamma$ which is more than one order of magnitude higher than an 
envelope made of heavy elements.

%-----------------------------------------------------------------------
%-----------------------------------------------------------------------

The chemical composition of this envelope is probably determined by 
poorly understood processes occurring during the first hours of the life
of the star, including post-supernova fall-back and also possible later
accretion, bombardment by high energy $\gamma$-rays from the 
magnetosphere, ejection of light nuclei by the pulsar mechanism,...
It is hence possibly totally unrelated to the interior structure of the 
star and may vary from star to star and/or evolve with time.
We have no choice but consider it as a free parameter which has to be 
varied independently of the internal structure of the star, i.e., 
within both the Minimal Scenario and any other exotic one.
Spectral fits to the thermal spectrum could determine the composition of
the atmosphere:
an iron atmosphere necessarily implies an heavy element envelope but a 
light element atmosphere unfortunately does not impose any restriction 
on the the chemical composition of the  layers a few tens of 
centimeters beneath it.

\vspace{-0.05in}

%%%%%%%%%%%%%%%%%%%%%%%%%%%%%%%%%%%%%%%%%%%%%%%%%%%%%%%%%%%%%%%%%%%%%%%%
\subsection{The neutrino luminosity}

The second important term in Eq.~\ref{Eq:energy-conservation} is $L_\nu$
which is strongly dominated by the neutrino emission from the core.
All significant processes are directly related to $\beta$- and inverse
$\beta$-decay of neutrons with protons and several of them are listed
in Table~\ref{Tab:nu} with their emissivities $q_\nu$.
The simplest such process is the direct Urca (``DUrca'') process.
However, momentum conservation in this process requires proton
fractions $x_p$ above 15\% [\refcite{LPPH91}] while at nuclear density 
it is only of the order of 5\%.
Thus, in the outer core of the neutron star, and this is the definition
of the outer core, neutrino emission is due to the modified Urca (MUrca)
process in which a second ``spectator'' nucleon (a neutron for the 
neutron branch or a proton in the proton branch of MUrca) contributes
by giving or absorbing the extra momentum needed.
Being a five fermion process instead of a three fermion one, the 
MUrca process is much less efficient than the DUrca process.
It acquires two extra Pauli blocking actors $(T/E_F)$,
$E_F$ being the Fermi energy of the extra nucleon:
since $E_F \sim$ 100 MeV, with $T=10^9\cdot T_9$ K, the reduction of 
MUrca is of the order of $10^{-6} \; T_9^2$ compared to DUrca.
Another possibility which allows the DUrca process, but with a reduced 
efficiency, is the presence of a charged meson ($\pi^-$ or $K^-$)
condensate which can easily contributes to momentum conservation
without introducing any dramatic phase space limitation as a
nucleon does in the MUrca process.
In case hyperons, or quarks, appear at high density they will
also participate in DUrca processes and enormously increase $L_\nu$.

In short, for the chemical composition expected at densities not too 
much higher than $\rho_{\rm nucl}$ where the proton fraction is small 
the neutrino emission is due to the MUrca process while any change 
beyond this will increase the emissivity by many orders of magnitude.
This MUrca process is the essence of the Standard Scenario for neutron 
star cooling but the occurrence of nucleon pairing and its proper 
treatment makes the subject more complicated and leads to the Minimal 
Scenario.
I refer the reader to the two excellent reviews by 
Pethick [\refcite{P92}] and Yakovlev et al. [\refcite{YKGH01}] for more
details on neutrino emission processes.

%-----------------------------------------------------------------------
\begin{table}[th]
\begin{center} 
\tbl{Some core neutrino emission processes and their emissivities.}
{\footnotesize
\begin{tabular}{ccc}
\hline %----------------------------------------------------------------
 Process Name & Process & $\begin{array}{c} {\rm Emissivity \;} q_\nu \\
                                            {\rm (erg/sec/cm^3)}
                           \end{array}$  \\
\hline %----------------------------------------------------------------
a) Modified Urca 
&  
$ \left\{ \begin{array}{c}
                 n'+n \rightarrow n'+p+e^-+\overline{\nu}_e \\ 
                 n'+p+e^- \rightarrow n'+n+\nu_e  
          \end{array} \right. $
&
$\sim 10^{21} \cdot T_9^8$  \\
%-----------------------------------------------------------------------
b) K-condensate 
&
$ \left\{ \begin{array}{c}
                 n+K^- \rightarrow n+e^-+\overline{\nu}_e \\   
                 n+e^- \rightarrow n+K^-+\nu_e
          \end{array} \right. $
& 
$\sim 10^{24} \cdot T_9^6$ \\
%-----------------------------------------------------------------------
c) $\pi$ - condensate 
&
$ \left\{ \begin{array}{c}
                 n+\pi^- \rightarrow n+e^-+\overline{\nu}_e \\
                 n+e^-   \rightarrow n+\pi^-+\nu_e
          \end{array} \right. $
&
$\sim 10^{26} \cdot T_9^6$ \\
%-----------------------------------------------------------------------
d) Direct Urca 
&
$ \left\{ \begin{array}{c}
                 n  \rightarrow p+e^-+\overline{\nu}_e \\
                 p+e^- \rightarrow n+\nu_e
          \end{array} \right. $
&
$\sim 10^{27} \cdot T_9^6$ \\
%-----------------------------------------------------------------------
e) Quark Urca 
&
$ \left\{ \begin{array}{c}
                 d \rightarrow u+e^-+\overline{\nu}_e \\
                 u+e^- \rightarrow d+\nu_e
          \end{array} \right. $
&
$\sim 10^{26} \alpha_c T_9^6$ \\
\hline %----------------------------------------------------------------
\end{tabular} \label{Tab:nu} }
\end{center}
\end{table}
%-----------------------------------------------------------------------

\vspace{-0.3in}

%%%%%%%%%%%%%%%%%%%%%%%%%%%%%%%%%%%%%%%%%%%%%%%%%%%%%%%%%%%%%%%%%%%%%%%%
\subsection{Baryon paring
            \label{Sec:Pairing}}

%-----------------------------------------------------------------------
\begin{figure}[b]
   \centerline{\epsfxsize=4.1in\epsfbox{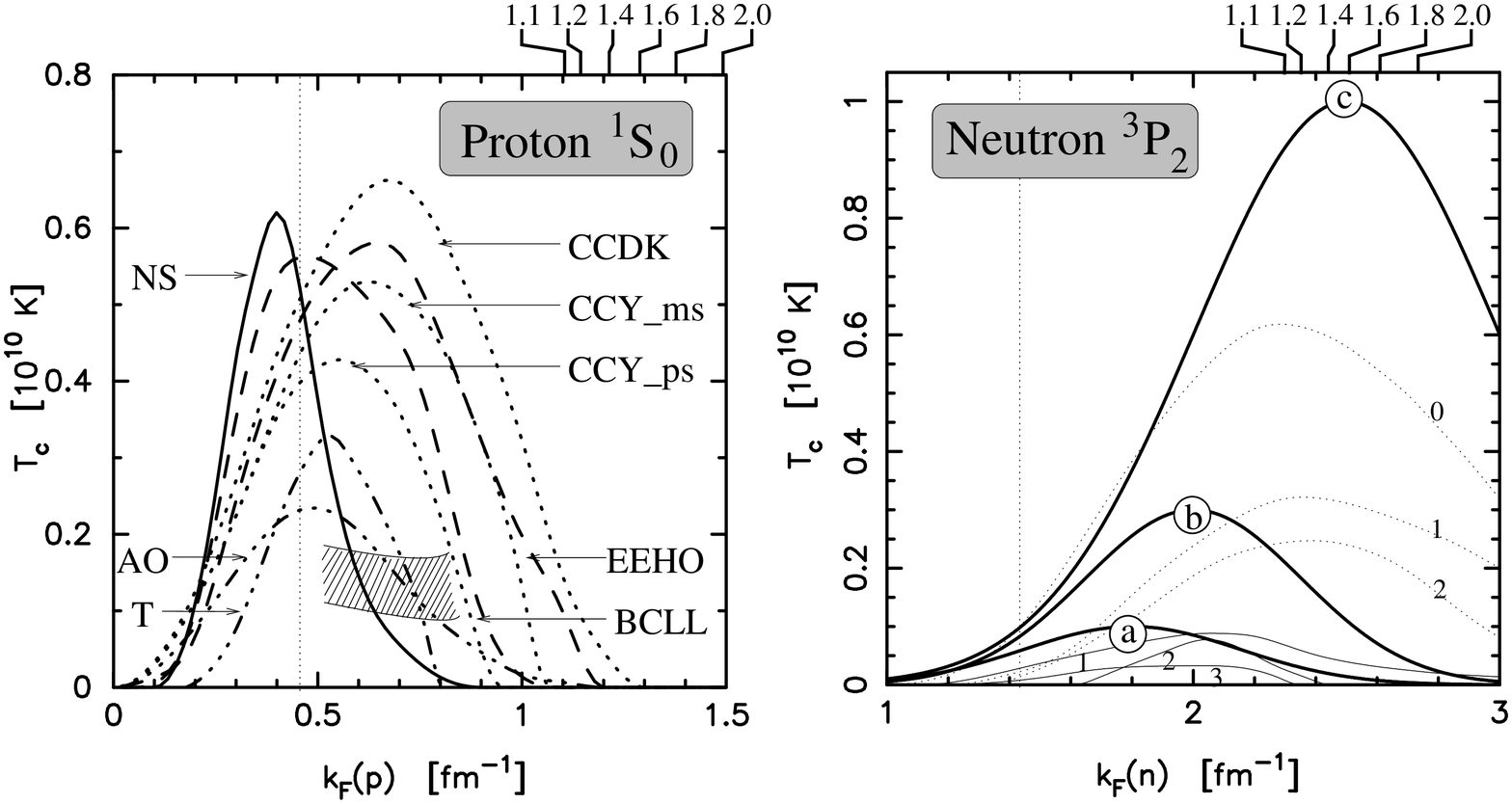}}   
   \caption{Predictions of critical temperatures $T_c$ for pairing of 
            protons in the
            $^1S_0$ state and neutrons in the $^3P_2$ state.
            The dotted vertical lines indicate the crust-core boundary.
            Values of $k_F$ at center of stars of masses 1.1, 1.2, 1.4,
            1.6, 1.8, and 2.0 \protect\Msun are marked at the upper 
            margin, for stars built with
            the EOS from [\protect\refcite{APR98}].
            See [\protect\refcite{PLPS04}] for references.
   \label{Fig:Tc} }
\end{figure}
%-----------------------------------------------------------------------

Pairing of baryons, either nucleons or hyperons, and also of quarks if 
present, is predicted to occur in most of the interior of a neutron 
star.
At low Fermi momenta neutrons and protons are expected to pair in a 
$^1S_0$ angular momentum state while at higher momenta a $^3P_2$ state 
is probably replacing it. 
The $^1S_0$ neutron gap has been extensively studied and is covering the
inner part of the crust with some extension in the outermost layers of 
the core.
The proton $^1S_0$ gap is also certainly present in the outer core and 
may or may not reach the center of the star, depending on the specific 
pairing model considered and on the central density of the star.
Figure~\ref{Fig:Tc} shows a representative sample of theoretical 
predictions for the associated critical temperature $T_c$.
Neutron pairing in the $^3P_2$ state is much more delicate and there is
a very wide range of predictions as is illustrated by the examples 
shown in Figure~\ref{Fig:Tc}.
As shown by Baldo et al [\refcite{BEEHS98}] the poor understanding of 
the nucleon-nucleon interaction in the $^3P_2$ state {\it in vacuum} 
by itself results in a wide range of predictions for $T_c$, illustrated
by the three curves labeled ``a'', ``b'', and ``c'' in 
Figure~\ref{Fig:Tc}.
Moreover, {\it in medium} effects were recently shown to have a 
dramatic effect on this gap [\refcite{SF04}] which may result to be 
vanishingly small.

The dramatic effect of pairing on the cooling comes from the gap it 
introduces in the single particle excitation spectrum which results in 
a strong suppression of both the specific heat and the neutrino 
emissivity of the paired component.
When $T \ll T_c$ this suppression is similar to a Boltzmann factor
$\exp(-\Delta/kT)$ and in general it is taken into account accurately by
multiplying the relevant $c_v$'s and $q_\nu$'s by appropriate 
``control functions'' (See Fig.~\ref{Fig:Pairing-effects}).

%%%%%%%%%%%%%%%%%%%%%%%%%%%%%%%%%%%%%%%%%%%%%%%%%%%%%%%%%%%%%%%%%%%%%%%%
\subsection{The Pair Breaking and Formation (``PBF'') 
            neutrino emission process
            \label{Sec:PBF}}

\begin{figure}[ht]
   \centerline{\epsfxsize=1.8in\epsfbox{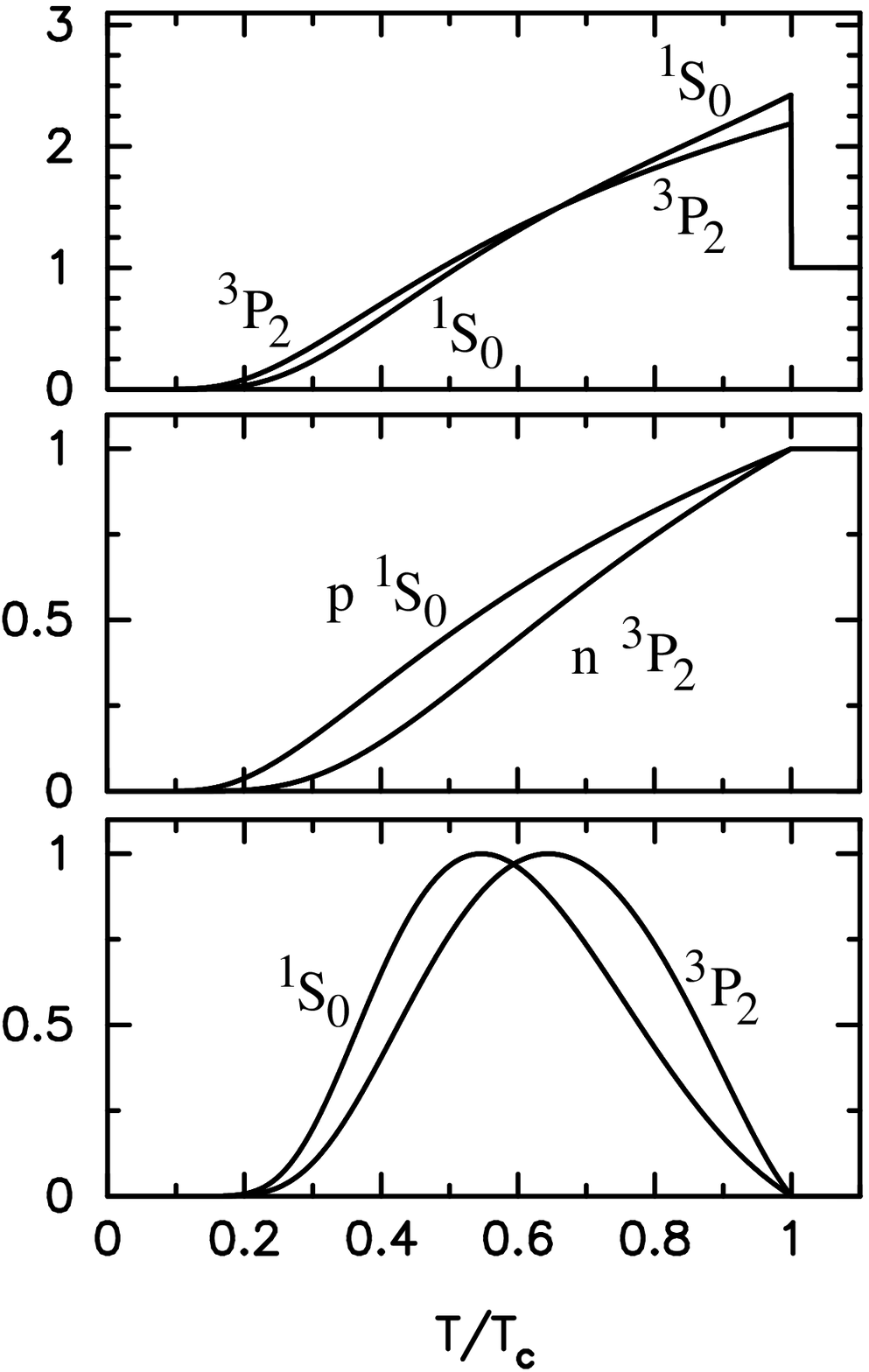} 
               \hspace{0.2in}
               \epsfxsize=2.2in\epsfbox{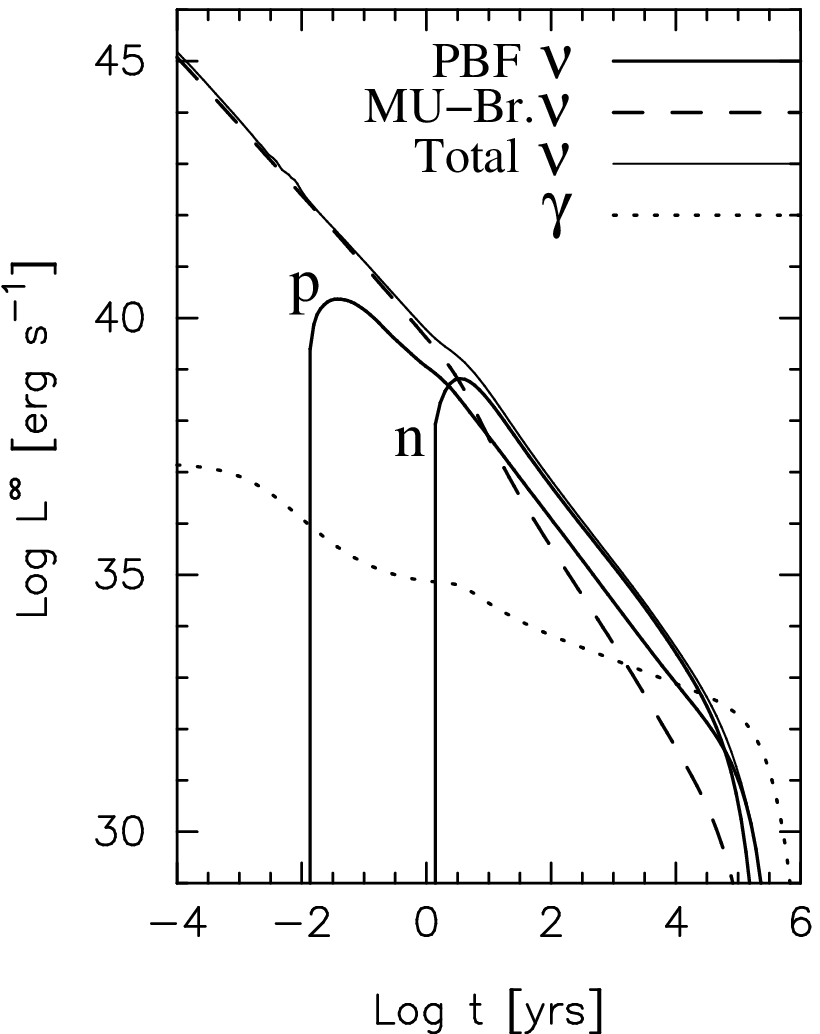}}
\caption{Left panel: 
         control functions, for pairing in the $^1S_0$ and $^3P_2$
         phases, of the specific heat (top), MUrca process (middle) and
         PBF process (bottom).
         Right panel: 
         comparison of the neutrino luminosities from the two PBF 
         processes due to neutron $^3P_2$ and proton $^1S_0$ gap  (the
         neutron $^1S_0$ gap contribution is small and not shown here),
         and from the MUrca processes, with the total neutrino
         luminosity and
         the photon luminosity (neutron $^3P_2$ gap from model ``a''
         and proton $^1S_0$ pairing from model AO, as labeled in
         Fig~\protect\ref{Fig:Tc}).
      \label{Fig:Pairing-effects}}
\end{figure}

The occurrence of pairing has a third effect, beside the suppression of
$c_v$ and $q_\nu$, which is the emission of $\nu-\bar{\nu}$ pairs
at temperature below, but close to, $T_c$ produced by the formation and
breaking of Cooper pairs, the ``PBF'' process [\refcite{FRS76,VS87}].
This process leads to a sudden increase of the neutrino emission in a
given layer, when $T$ reaches $T_c$, which can largely dominates over 
the emission from the MUrca process.
For example, in the case of the neutron $^3P_2$ pairing its emissivity 
is
\be
q_\nu^{\rm n ^3P_2} =
   8.6 \times 10^{21} \left(\frac{\rho}{\rho_0}\right)^{1/3}
   \left(\frac{m_n^*}{m_n}\right)  \times
   \tilde{F}_{\rm ^3P_2}(T/T_c) 
   \left(\frac{T}{10^9 \rm K}\right)^7
\ee
The control functions $\tilde{F}$ are plotted in 
Fig.~\ref{Fig:Pairing-effects}
and describe the onset of the process when $T$ reaches $T_c$ and its
suppression when $T \ll T_c$.
Comparing the emissivities given above with the MUrca process in 
Table~\ref{Tab:nu} one sees that these PBF processes can dominate the 
neutrino emission and we will see in the next section that they are an 
essential ingredient of the Minimal Scenario.

%-----------------------------------------------------------------------
%----------------------------------------------------------------------

%\vspace{-0.3in}

%%%%%%%%%%%%%%%%%%%%%%%%%%%%%%%%%%%%%%%%%%%%%%%%%%%%%%%%%%%%%%%%%%%%%%%%
\subsection{The specific heat}

For normal (i.e., unpaired) degenerate spin $\frac 12$ fermions
of type ``i'', the specific heat (per unit volume) is
\be
c_{i,v} = N_{i}(0) \frac{\pi^2}{3} k_B^2 T 
          = \frac{m_i^* n_i}{p_{i,F}^2}  \pi^2 k_B^2 T 
\label{Eq:Cv_deg}
\ee
Most of the specific heat of the star is provided by the core and,
in absence of ``exotic'' matter, nucleons contribute about 90\% of it 
while leptons ($e$ and $\mu$) share the remaining.
Once neutrons and/or protons go into a paired state their specific heat 
is strongly altered: 
when $T$ reaches $T_c$ there is a discontinuity in $c_v$ which suddenly
increases but when $T \ll T_c$ it becomes exponentially suppressed.
This effect is also accurately taken into account by introducing a
multiplicative ``control function'' plotted in 
Fig.~\ref{Fig:Pairing-effects}.
It is important to notice that even in case both neutrons and protons 
are paired in the whole core we still have the contribution of the 
leptons which remains untouched,
i.e., pairing can reduce the total $C_v$ by at most 90\% while $L_\nu$ 
can be suppressed by many orders of magnitude in case baryons involved 
in all the important processes are paired.

%%%%%%%%%%%%%%%%%%%%%%%%%%%%%%%%%%%%%%%%%%%%%%%%%%%%%%%%%%%%%%%%%%%%%%%%
%%%%%%%%%%%%%%%%%%%%%%%%%%%%%%%%%%%%%%%%%%%%%%%%%%%%%%%%%%%%%%%%%%%%%%%%
\section{The Minimal Scenario
         \label{Sec:Minimal}}
%%%%%%%%%%%%%%%%%%%%%%%%%%%%%%%%%%%%%%%%%%%%%%%%%%%%%%%%%%%%%%%%%%%%%%%%
%%%%%%%%%%%%%%%%%%%%%%%%%%%%%%%%%%%%%%%%%%%%%%%%%%%%%%%%%%%%%%%%%%%%%%%%

The physical ingredients presented in the previous section constitute
all the essential input for the Minimal Scenario.
Since, by definition, this scenario does not admit any enhanced neutrino
emission the cooling history of a neutron star has only a very weak
dependence on its mass.
Moreover, the supranuclear EOS is also well constrained within this
scenario so that we can generally simply study the evolution of a 
``canonical'' 1.4 \Msun neutron star.
All results presented here are based on the EOS from APR 
[\refcite{APR98}].
What is {\em not} constrained by the requirement of the 
Minimal Scenario is:

\smallskip
\noindent
{\bf A) the chemical composition of the envelope} and \\
{\bf B) the pairing state of the nucleons}

\smallskip
\noindent
and the large uncertainties in these two physical ingredients are, by 
far, the most important sources of uncertainty in the theoretical 
predictions of the Minimal Scenario.

\begin{figure}[ht]
   \centerline{\epsfxsize=2.00in\epsfbox{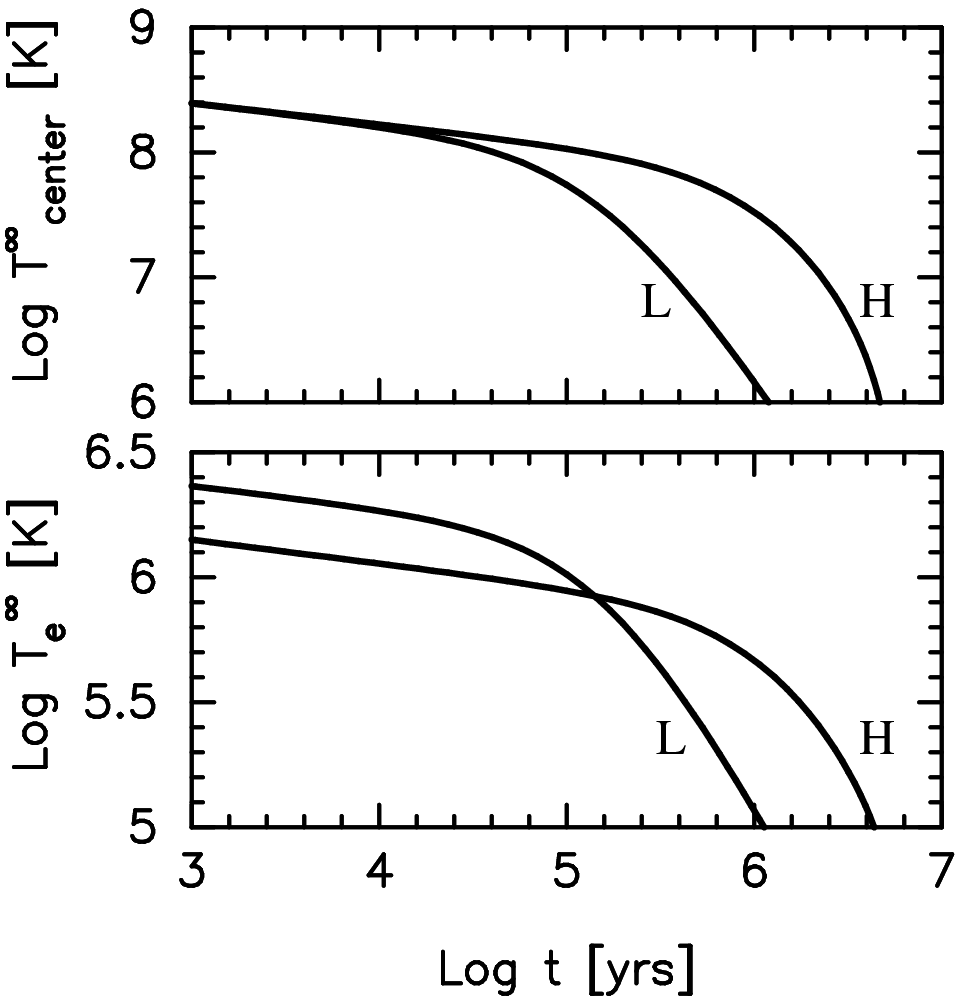} 
               \hspace{0.2in} 
               \epsfxsize=2.15in\epsfbox{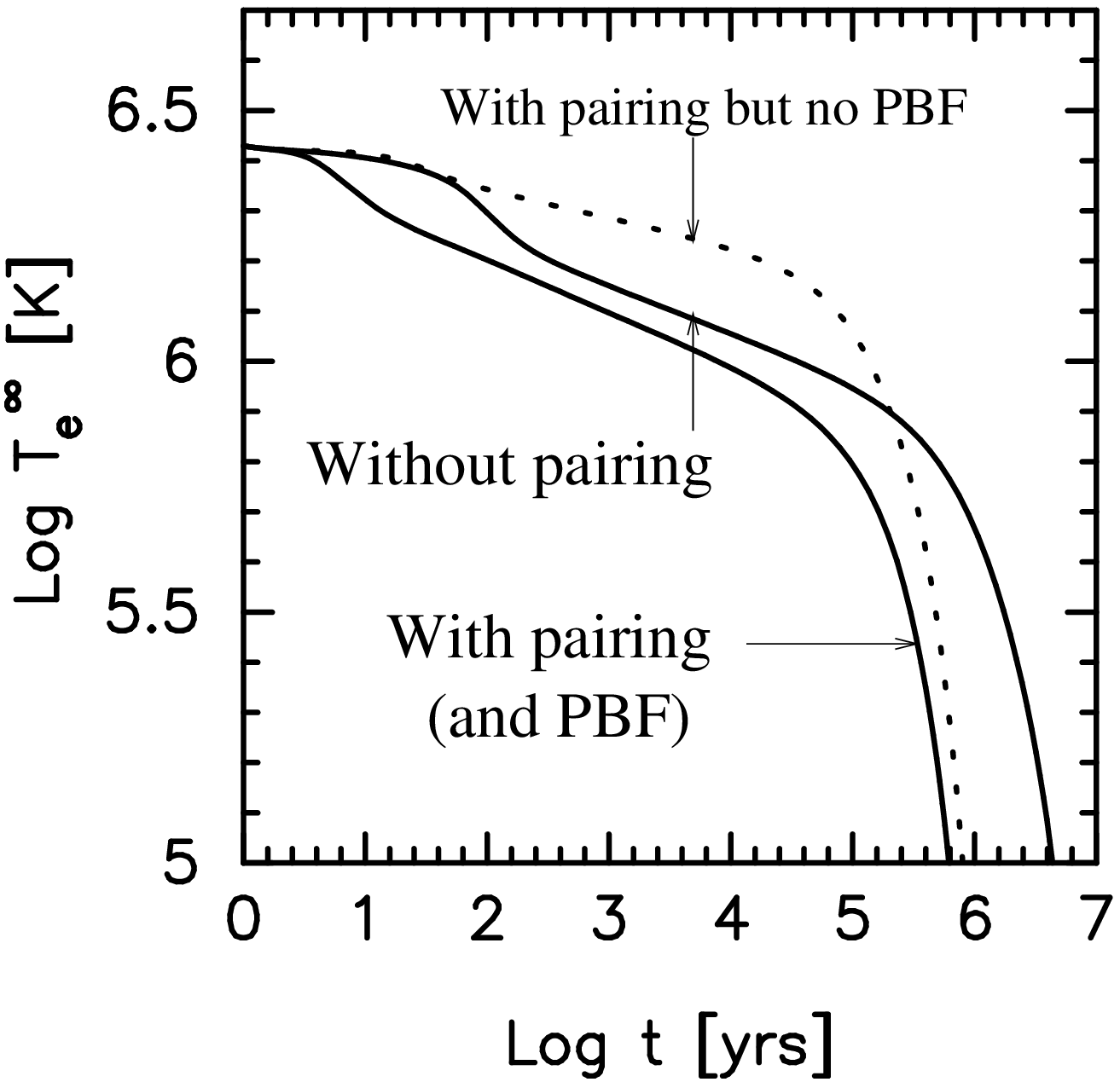}}
\caption{Left panel: 
         effect of the envelope chemical composition in the cooling.
         Right panel: 
         effect of nucleon pairing on the cooling.
         (See text for description.)
         \label{Fig:Trivial}}
\end{figure}

The effect of the envelope is illustrated in the left panel of
Fig.~\ref{Fig:Trivial} for the two extreme cases of a star with an 
envelope consisting only of heavy elements (marked as ``H'') and with 
an envelope containing a maximum amount of light elements 
(marked as ``L'').
The important features to note are: \\
{\bf A1)} first, at age inferior to $10^5$ yrs,
both stars have the same central temperature but the ``L'' model has a 
higher $T_e$: 
this correspond to the {\em neutrino cooling era} where $L_\nu$ 
drives the cooling, hence the same $T_{\rm center}$ for both stars,
and the surface temperature simply follows the interior evolutions, 
hence a higher $T_e$ in presence of a less insulating light element 
envelope, and \\
{\bf A2)} later, during the {\em photon cooling era} when 
$L_\gamma \gg L_\nu$,  the cooling trajectories get inverted since the 
light element envelope results in a much larger photon luminosity and 
hence faster cooling. \\
With an envelope containing a smaller amount of light elements we obtain
an intermediate evolution and in case we allow for a time evolution of 
the amount of light elements the evolution can switch from one 
trajectory to the other.

The overall effect of pairing is illustrated in the three cooling curve
plotted in the right panel of Fig.~\ref{Fig:Trivial}.
The two important features to note are: \\
{\bf B1)} comparing the model without pairing with the model with 
pairing but without the PBF process taken into account, one sees the 
effect of the suppression of the MUrca neutrino process resulting in a 
warmer star during the neutrino cooling era while during the photon 
cooling era the results are inverted because of the suppression of 
$C_v$ from the pairing, and \\
{\bf B2)} once the neutrino emission from the PBF process is taken into
account the cooling is strongly enhanced during the neutrino cooling 
era, confirming the results of Fig~\ref{Fig:Pairing-effects} (right 
panel) that this PBF process can be much more intense than the MUrca 
one, and finally during the photon cooling era the two paired models, 
with and without PBF, join once they have forgotten their previous 
neutrino cooling history. \\
Of course the PBF process is always acting in presence of pairing and it
has been artificially turned-off for this figure, but its efficiency 
depends on the actual size of the gap, i.e., the actual profile of 
$T_c$ for either neutrons or protons, and the gaps used in 
Fig~\ref{Fig:Pairing-effects} have been chosen to maximize the effect.

%-----------------------------------------------------------------------
%----------------------------------------------------------------------

\begin{figure}
   \centerline{\epsfxsize=4.4in\epsfbox{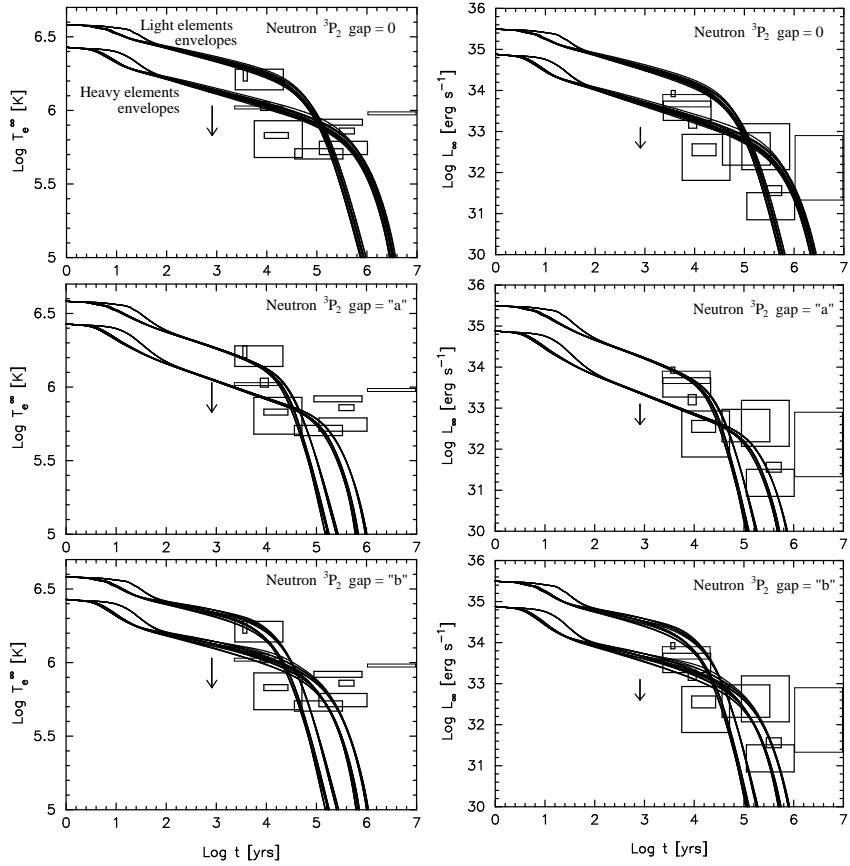}}
   \caption{Comparison of predictions of the Minimal Scenario with data.
            Left panels: 
            effective temperature at infinity $T_e^\infty$ vs. age.
            Right panels: 
            luminosity at infinity $L_\infty$ vs age.
            The upper, middle, and lower panels correspond to three 
            different assumption about the size of the neutron $^3P_2$ 
            gap as indicated in the panels.
         In each panel the two sets of curves correspond to the two
         extreme models of envelope chemical composition: 
         light elements or heavy elements, as labeled in the upper left
         panel.
         For each set of curves, the different 15 curves correspond to 
         different choices of the neutron (3 cases) and proton (5 cases)
         $^1S_0$ gaps.
         1.4 \Msun star built with the EOS of APR.
         \label{Fig:plot_cool_lum_data}}
\end{figure}

An extensive comparison of the predictions of the Minimal Scenario with
the data presented in Sec.~\ref{Sec:Data} is shown in 
Fig~\ref{Fig:plot_cool_lum_data}.
For the reasons discussed in Sec.~\ref{Sec:Data} results are plotted 
twice, as $T_e$ vs age (left panels) and $L$ vs age (right panels).
The uncertainties due to the exact extent of nucleon pairing are better
assessed by classifying the possible models into three families 
depending on the size of the neutron $^3P_2$ gap since this is the most
uncertain one:
a vanishing gap and the schematic models ``a'' and ``b'' of 
Fig.~\ref{Fig:Tc}.
For each of these three families variations of the neutron and proton 
$^1S_0$ gaps generate a set of closely packed curves and, given the 
uncertainty about the envelope chemical composition each set is shown 
twice, assuming an envelope made of heavy elements and an envelope with
a maximum amount of light elements.
(For clarity only these two extreme cases of envelope are shown,
but any trajectory inbetween is possible.)

%-----------------------------------------------------------------------
%-----------------------------------------------------------------------

The overall agreement between theory and data is quite good, which I 
personally find in itself amazing considering that this is REAL THEORY:
these calculations culminate several decades of works from hundreds
of physicists and astrophysicists based essentially on only a handful of
observational facts (the very existence of ``neutron stars'', several 
mass measurements and their extreme compactness known from pulsar 
timing, ...).

The three sets of models, for the three different neutron $^3P_2$ gaps,
are quite similar but do show some essential differences.
When considering young stars, particularly J0205+6449 (in 3C58),
PSR 0833-45 (in Vela) and PSR 1706-44, one sees that the models with 
the $^3P_2$ gap ``a'' are very close to the upper limits of Vela and 
3C58 while the difference is larger with the other two gaps.
The interpretation of the data of PSR 1706-44 is more ambiguous due to 
the presently large uncertainty on both it temperature (or luminosity) 
and age.
Since no thermal emission has been actually detected from 3C58 it is 
more prudent to consider it on a $L$-age plotted where the discrepancy 
with the theoretical predictions is actually the largest.

Several of the older objects may have temperatures higher than some of 
the theoretical predictions of the Minimal Scenario.
This may be attributed to an erroneous age, considering that the only 
information we have bout their possible age is the spin-down time scale
which can be very misleading.
Another possibility is that some ``heating mechanism' is at work which
converts rotational, or magnetic, energy into heat.

%%%%%%%%%%%%%%%%%%%%%%%%%%%%%%%%%%%%%%%%%%%%%%%%%%%%%%%%%%%%%%%%%%%%%%%%
%%%%%%%%%%%%%%%%%%%%%%%%%%%%%%%%%%%%%%%%%%%%%%%%%%%%%%%%%%%%%%%%%%%%%%%%
\section{Conclusions}
         \label{Sec:Conclusion}
%%%%%%%%%%%%%%%%%%%%%%%%%%%%%%%%%%%%%%%%%%%%%%%%%%%%%%%%%%%%%%%%%%%%%%%%
%%%%%%%%%%%%%%%%%%%%%%%%%%%%%%%%%%%%%%%%%%%%%%%%%%%%%%%%%%%%%%%%%%%%%%%%

In the Quest for New States of Dense Matter we have performed an 
extensive study of the thermal evolution of isolated neutron stars under
the hypothesis that no new phase is present and tried to find some 
incompatibility of the results of this assumption with the best 
presently available data on cooling neutron stars.
The final results, presented in Fig.~\ref{Fig:plot_cool_lum_data}
show now striking incompatibility with, nevertheless two objects,
J0205+6449 (in 3C58) and PSR 0833-45, which are conspicuously lower
than our predictions. 
Given the capability of both {\em Chandra} and {\em XMM-Newton}
one can have the hope that in the near future either more such objects 
will be found (see, e.g., Kaplan {\it et al.} [\refcite{Ketal}])
or that more data on these two conspicuous
stars will permit more detailed studies and confirm them as
star(s) which encompass a new state of dense matter.

%%%%%%%%%%%%%%%%%%%%%%%%%%%%%%%%%%%%%%%%%%%%%%%%%%%%%%%%%%%%%%%%%%%%%%%%
\section*{Acknowledgments}
Work partially supported by grants from UNAM-DGAPA (\#IN112502) and 
Conacyt (\#36632-E).
The author also wants to thank the organizers of this Workshop for
the invitation and financial support.

%%%%%%%%%%%%%%%%%%%%%%%%%%%%%%%%%%%%%%%%%%%%%%%%%%%%%%%%%%%%%%%%%%%%%%%%
%%%%%%%%%%%%%%%%%%%%%%%%%%%%%%%%%%%%%%%%%%%%%%%%%%%%%%%%%%%%%%%%%%%%%%%%

%%%%%%%%%%%%%%%%%%%%%%%%%%%%%%%%%%%%%%%%%%%%%%%%%%%%%%%%%%%%%%%%%%%%%%%%
%%%%%%%%%%%%%%%%%%%%%%%%%%%%%%%%%%%%%%%%%%%%%%%%%%%%%%%%%%%%%%%%%%%%%%%%

%%%%%%%%%%%%%%%%%%%%%%%%%%%%%%%%%%%%%%%%%%%%%%%%%%%%%%%%%%%%%%%%%%%%%%%%
%%%%%%%%%%%%%%%%%%%%%%%%%%%%%%%%%%%%%%%%%%%%%%%%%%%%%%%%%%%%%%%%%%%%%%%%
%%%%%%%%%%%%%%%%%%%%%%%%%%%%%%%%%%%%%%%%%%%%%%%%%%%%%%%%%%%%%%%%%%%%%%%%
\end{document}